\documentclass[cmp]{svjour}

\usepackage{amsmath,amsfonts,amssymb,graphics,graphicx,epsfig,color,times,bbm}

\newcommand{\rr}{\mathbbm{R}}

\newcommand{\id}{\mathbbm{1}}

\begin{document}

\newcommand{\proofend}{\hfill\rule{2mm}{2mm} \\}

\title{Gaussian quantum marginal problem}

\author{Jens Eisert\inst{1,2}, Tom\'a\v{s} Tyc\inst{3}, Terry Rudolph\inst{2}, and Barry C.\ Sanders\inst{4}}

\institute{Institute for Mathematical Sciences, Imperial College London, 
	Princes Gate, London SW7 2PE, UK
		\and
	Blackett Laboratory, Imperial College London, Prince Consort Road, London SW7 2BW, UK
		\and
	Institute of Theoretical Physics, Masaryk University, 61137 Brno, Czech Republic
		\and
	Institute for Quantum Information Science, University of Calgary, Alberta T2N 1N4, Canada}
\date{\today}

\maketitle

\begin{abstract}
The quantum marginal problem asks what local spectra are consistent with a given spectrum of a joint 
state of a composite quantum system. This setting, also referred to as the question of the compatibility of local spectra, has several applications in quantum information theory. Here, we
introduce the analogue of this statement for Gaussian states for any number of modes, and solve it in
generality, for pure and mixed states,
both concerning necessary and sufficient conditions.
Formally, our result can be viewed as an
analogue of the Sing-Thompson Theorem
(respectively Horn's Lemma), characterizing the
relationship between main diagonal elements and
singular values of a complex matrix: We
find necessary and sufficient conditions
for vectors $(d_1,\dots, d_n)$ and $(c_1,\dots, c_n)$
to be the symplectic eigenvalues
and symplectic main diagonal elements
of a strictly positive real matrix,
respectively. More physically speaking, this result
determines what
local temperatures or entropies are consistent with a
pure or mixed Gaussian state of several modes. We find
that this result implies a solution to the problem of 
sharing of entanglement in pure Gaussian states
and allows for estimating the global entropy of non-Gaussian states
based on local measurements.
Implications to the actual preparation of
multi-mode continuous-variable entangled states are discussed.
We compare the findings with the
marginal problem for qubits, the solution of which for
pure states has a strikingly similar and in fact simple
form.
\end{abstract}

\maketitle

\section{Introduction}

What reduced states are compatible with a 
quantum state of a composite system? The study of
this question has in fact a long tradition -- as the natural
quantum analogue of the marginal problem in
classical probability theory. Very recently,
this problem, now coined the
{\it quantum marginal problem},
has seen a revival of interest,
motivated by applications in the context of quantum information
theory \cite{Higuchi,Higuchi2,Bravyi,Han,Discrete,Franz}.
In fact, in the quantum information setting,
notably in quantum channel capacity expressions,
in assessments
of quantum communication protocols, or in the separability
problem, one often encounters
questions of compatibility of reductions with global quantum states
\cite{Squashed,THL,NK,JE,Daftuar}.

Since it is only natural to look at the full orbit under local unitary
operations, the quantum marginal problem immediately translates to a
question of the compatibility of spectra of quantum states. The
{\it mixed quantum marginal problem} then amounts to the following question:
Is there a state $\rho$ of a quantum system with $n$ subsystems, each
with a reduction $\rho_k$, that is
consistent with
\begin{eqnarray}
    \text{spec} (\rho) &=&r,\\
    \text{spec} (\rho_k) &= & r_k
\end{eqnarray}
for $k=1,\dots, n$, $r$ and $r_k$ denoting the respective
vectors of spectra.
In the {\it pure marginal problem}, one assumes $\rho=|\psi\rangle\langle\psi|$
to be pure.
In the condensed-matter context \cite{Hall,NRep}, 
related questions are also of
interest: For example, once one had classified all possible two-qubit
reductions of translationally invariant quantum states,
then one would
be able to obtain the ground state energy of any nearest-neighbor
Hamiltonian of a spin chain. The quantum marginal problem
was solved in several steps:
Higuchi et al. \cite{Higuchi}
solved the pure quantum marginal problem for qubits.
Subsequently, Bravyi was able to solve the mixed state case
for two qubits, followed
by Franz \cite{Franz} and  Higuchi \cite{Higuchi2}
for a three qutrit system. The general solution of the
quantum marginal problem for finite-dimensional systems was found
in the celebrated work of Klyachko \cite{Discrete}, see also Refs.\
\cite{Christandl,PhDC}. This is indeed
a closed-form solution. Yet the number of constraints grows
extremely rapidly with the system size, rendering the
explicit check whether the conditions are satisfied
unfeasible even for relatively small systems.

In this work, we introduce the Gaussian version of the
quantum marginal problem. Gaussian states play a
key role in a number of contexts, specifically
whenever bosonic modes and quadratic Hamiltonians become relevant, which are ubiquitous in
quantum optical systems,
free fields, and condensed matter lattice systems. For general infinite dimensional systems the marginals problem may well be intractable. However,
given that in turn these Gaussian states can be described by merely their
first and second moments \cite{Survey,Peter},
one could reasonably hope that it could be
possible to give a full account
of the {\it Gaussian quantum marginal problem}.
This gives rise, naturally, not to a condition to spectra of
quantum states, but to symplectic spectra, as explained
below. For the specific case of  three modes, the result is
known \cite{Alessio}, see also Ref.\ \cite{Rev}.
In this work we will show that this program of
characterizing the reductions of Gaussian states
can be achieved in generality, even concerning both
necessary and sufficient conditions. This means that one
can give a complete answer to what reductions entangled
Gaussian states can possible have.\footnote{We refer here to 
the marginal problem for Gaussian states, which are quantum
states fully defined by their first and second moments of canonical
coordinates. However, clearly, our result equally applies to general
and hence non-Gaussian states, in that it fully answers the question
what local second moments are consistent with global second 
moments of quantum states of several modes.}

Equivalently, we can describe this Gaussian marginal
problem as a problem of compatibility of temperatures
of standard harmonic systems: Given a state $\rho$, what
{\it local temperatures} -- or equivalently for single modes, what
{\it local entropies} -- are compatible with this
joint state? Of course, one can always take the temperatures to
be equal. But if they are different, they constrain each other in
a fairly
subtle way, as we will see. In a sense, the result gives rise to the
interesting situation that by looking at local temperatures, one can assess whether these reductions
may possibly originate from a joint system in a
pure state. Finally, it is important to note, since sufficiency of the conditions is always proven by an
explicit construction, the result also implies a recipe for
{\it preparing multi-mode continuous-variable entangled
states}.

\begin{figure}
  \includegraphics[width=8.5cm]{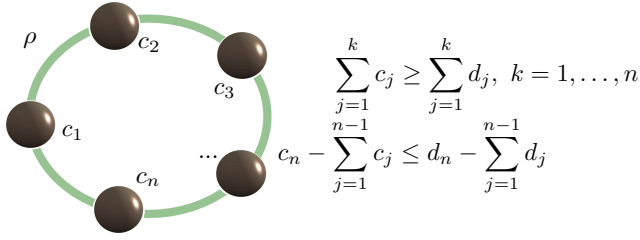}
  \caption{Solution of the Gaussian marginal problem.
  The set of possible reductions with symplectic spectra
  $(c_1,\dots, c_n)$ of a correlated or entangled
  Gaussian state $\rho$ with
  symplectic spectrum $(d_1,\dots, d_n)$ is characterized
  by the given remarkably simple necessary and sufficient set of
  $n+1$ inequalities. For example,
  from local measurements, one can hence
  infer about the consistency with the
  purity of the joint state. It also governs the sharing of
  correlations in Gaussian states.}
\end{figure}

\section{Main result}

 We consider states on $n$
modes, and consider reductions to single modes. Gaussian states are
represented by the matrix of second moments, the
$2n\times 2n$ covariance matrix $\gamma$
of the system, together with the vector $\mu $
of first moments. For a definition and a survey of properties,
see Refs.\ \cite{Survey,Peter}.
In this language, the vacuum state of a standard oscillator becomes
$\gamma = \id_2$, as the $2\times 2$ identity matrix. The canonical
commutation relations are embodied in the symplectic matrix
\begin{equation}\label{SymplecticMatrix}
    \sigma=\bigoplus_{k=1}^n
    \left[
    \begin{array}{cc}
    0 & 1\\
    -1 & 0\\
    \end{array}
    \right]
\end{equation}
for $n$ modes. The {\it covariance matrices} of $n$ modes
are exactly those real matrices satisfying
\begin{equation}
    \gamma + i\sigma\geq 0,
\end{equation}
which is simply a statement of the Heisenberg uncertainty 
principle. The first moments can always
be made zero locally, and are hence not
interesting for our purposes here. Note also that the
set of Gaussian states is closed under reductions,
so reduced states of Gaussian states are always
Gaussian as well.

Real matrices that leave the
symplectic form invariant, $S\sigma S^T = \sigma$,
form the real symplectic group Sp$(2n,\rr)$.
In the same way as symmetric matrices $M$
can be diagonalized
with orthogonal matrices to a diagonal matrix
$O M O^T =D$, one can diagonalize strictly
positive matrices using such $S\in \text{Sp}(2n,\rr)$,
according to
\begin{equation}
    SMS^T = D.
\end{equation}
The simply counted
main diagonal elements of $D$ form then the {\it symplectic
spectrum}
of $M$, and the collection of symplectic eigenvalues can be
abbreviated as $\text{sspec}(M)=(d_1,\dots, d_n)$,
\begin{equation}
    D=\text{diag} (d_1,d_1,\dots, d_n,d_n).
\end{equation}
This procedure is nothing but
the familiar {\it normal mode decomposition}.
In turn, by definition, the
symplectic eigenvalues are given by the square roots of the eigenvalues
of the matrix $-M\sigma M\sigma$. Again, for the vacuum, the symplectic
eigenvalues are all given by unity. In a mild abuse of notation,
we will refer to the symplectic spectrum of a Gaussian state
as the symplectic spectrum of the respective covariance matrix.

Finally, for a given covariance matrix $\gamma$, and in fact any
strictly positive real matrix, we refer to the
{\it symplectic main diagonal elements} $(c_1,\dots, c_n)$
as the symplectic eigenvalues of the $2\times 2$ main
diagonal blocks. This is the natural analogue of main diagonal
elements.
Equivalently, the symplectic main diagonal elements
are the main diagonal elements after the main diagonal
$2\times 2$ blocks have  been brought into the form
\begin{equation}\label{mdf}
    \gamma_k =\left[
    \begin{array}{cc}
    c_k & 0\\
    0 & c_k\end{array}
    \right].
\end{equation}
We are now in the position to state
our main result, see Fig.\ 1. 
It relates the symplectic spectrum of
composite systems to the ones of the reductions. We will first
state it as a mere matrix constraint, then as the actual
Gaussian marginal problem, and finally for the important
special case of having a pure joint state.

\begin{theorem}[Necessary and sufficient conditions] Let
$(d_1,\dots, d_n)$ and $(c_1,\dots, c_n)$ be two vectors
of positive numbers in non-decreasing
order. Then there exists a
strictly positive real $2n\times 2n$-matrix
$\gamma$ such that
$(d_1,\dots, d_n)$ are its symplectic eigenvalues and $(c_1,\dots, c_n)$ the
symplectic main diagonal elements
if and only if the $n+1$ conditions
    \begin{eqnarray}\label{C1}
        \sum_{j=1}^k c_j
        &\geq& \sum_{j=1}^k d_j ,\,k=1,\dots, n\\
    \label{C2}
        c_n  -  \sum_{j=1}^{n-1} c_j&\leq&
        d_n  -  \sum_{j=1}^{n-1} d_j
    \end{eqnarray}
    are satisfied.
\end{theorem}

This set of inequalities may be conceived as a general analogue of the
Sing-Thompson theorem \cite{Sing,Tho1,Tho}, see below.
More physically speaking, this means the following:

\begin{corollary}[Gaussian marginal problem]\label{MixedProblem}
    Assume that $\rho$ is a Gaussian state of $n$ modes satisfying
    $\text{sspec}(\rho) = (d_1,\dots, d_n)$.
    Then the possible reduced states $\rho_k$ to
    each of the individual modes $k=1,\dots, n$
    are exactly those Gaussian states with
    \begin{equation}
        \text{sspec}(\rho_k) = c_k
    \end{equation}
    satisfying Eq.\ (\ref{C1}) and (\ref{C2}).
\end{corollary}

These conditions hence fully characterize the possible reduced marginal states.
For two modes, $n=2$, for example, the given conditions read
\begin{eqnarray}
    c_1+c_2 &\geq& d_1+d_2,\\
    c_2-c_1&\leq & d_2-d_1,
\end{eqnarray}
for $c_2\geq c_1 $ and $d_2\geq d_1$. The constraint
$c_1\geq d_1$ is then automatically satisfied.
For pure Gaussian states, the above conditions 
take a specifically
simple form. Quite strikingly,
we will see that the resulting conditions very much resemble the situation
of the marginal problem for qubits.

\begin{corollary}[Pure Gaussian marginal problem]\label{pureprob}
    Let $\rho=|\psi\rangle\langle\psi|$
    be a    pure Gaussian state of $n$ modes.
    Then the set $(b_1+1,\dots, b_n+1)$
    of symplectic eigenvalues
    \begin{equation}
        \text{sspec}(\rho_k) = b_k+1
    \end{equation}
    $k=1,\dots, n$, of the reduced states $\rho_k$
    of each of the $n$ modes is given by the set
    defined by
    \begin{equation}\label{Cond}
        b_{j} \leq \sum_{k\neq j} b_k
    \end{equation}
        for all $j$, for $b_j\geq 0$.
\end{corollary}
To reiterate, these conditions are necessary and sufficient for
the local symplectic spectra being consistent with the global state
being a pure Gaussian state.

Equivalently, this can be put as follows:
If $\gamma$ is the covariance matrix of a pure Gaussian state
with reductions
    \begin{equation}\label{Form}
    \gamma_k=\left[
    \begin{array}{cc}
    b_k +1& 0 \\
    0 & b_k+1
    \end{array}
    \right],
\end{equation}
$k=1,\dots ,n$. Then, Eq.\ (\ref{Cond}) defines the
local temperatures $T_k$
per mode consistent with the whole system being in a
pure Gaussian state, according to
\begin{equation}
    b_k =2 (\exp(1/T_k)-1)^{-1},
\end{equation}
for the standard harmonic oscillator (an oscillator with unit mass and
frequency).
The above condition hence determines the {\it temperatures} that modes
can have, given that a composite system is in a pure Gaussian state.
The form
of Eq.\ (\ref{Form}) can always be achieved by means
of local rotations and squeezings in phase space. One can hence equally
think in terms of local symplectic spectra or local temperatures.

It is instructive to compare the results for the pure Gaussian
marginal problem with the one for {\it qubits} as solved in
Ref.\ \cite{Higuchi}. There, it has been found that for a
system consisting of $n$ qubits, one has
\begin{equation}
    \lambda_j \leq \sum_{k\neq j} \lambda_k
\end{equation}
for the spectral values $r_k=(\lambda_k,1-\lambda_k)$,
$\lambda_k\in [0,1]$. 
Moreover, these
conditions
are both necessary and sufficient. It is remarkable that this
form is identical with the result for $n$ single modes
\begin{equation}
        b_{j} \leq \sum_{k\neq j} b_k,
\end{equation}
$b_k\geq 0$,
as necessary and sufficient conditions.
Again,
the admissible symplectic eigenvalues are defined by a cone the base of 
which is formed by a simplex.
Note that the methods used in Ref.\ \cite{Higuchi} to arrive at
the above result are entirely different. Once again, a
striking formal similarity between the case of
qubit systems and Gaussian states is encountered.

Finally, from the perspective of matrix analysis,
the above result can be seen as a general
analogue of the {\it Sing-Thompson Theorem} \cite{Sing,Tho1,Tho}
(or {\it Horn's Lemma} \cite{Horn}
in case of Hermitian matrices), first posed in Ref.\
\cite{Mirsky}, where the role of
singular values is taken by the symplectic eigenvalues. \\

\noindent
{\bf Sing-Thompson Theorem (\cite{Sing,Tho1,Tho})}\label{Thompson}
{\it Let $(x_1,\dots, x_n)$ be complex
numbers such that $|x_k|$ are non-increasingly ordered
and let $(y_1,\dots, y_n)$ be non-increasingly ordered
positive numbers. Then an $n\times n$ matrix exists with
$x_1,\dots, x_n$ as its main diagonal and $y_1,\dots, y_n$
as its singular values if and only if
\begin{eqnarray}\label{T}
    \sum_{j=1}^k |x_j| &\leq&
    \sum_{j=1}^k y_j,\,\, k=1,\dots, n,\\
    \sum_{j=1}^{n-1}  |x_j| - |x_n|
     & \leq &  \sum_{j=1}^{n-1} y_j- y_{n}.
\end{eqnarray}
}
It is interesting to see that -- although the symplectic
group $\text{Sp}(2n,\rr)$
is not a compact group -- there is so much formal
similarity concerning the implications on main
diagonal elements of matrices. Note, however, that the ordering
of singular values
and symplectic eigenvalues, respectively,
is different in Theorem 1 and in the Sing-Thompson theorem.
\section{Proof}

As a preparation of the proof, we will identify
a simple set of necessary conditions that
constrains the possible reductions that are consistent with
the assumption that the state is pure and Gaussian.
These simple conditions derive from a connection between
the symplectic trace and the trace of the covariance matrix.
Quite surprisingly, we will see that they already define the full
set of possible marginals consistent with a Gaussian state
of $n$ modes. We shall start by stating the condition to the
reductions.

\begin{lemma}[Symplectic trace]\label{ST}
Let $\gamma$ be a strictly positive real $2n\times 2n$-matrix
such that its main diagonal $2\times 2$ blocks
are given by Eq.\ (\ref{mdf}) for $c_k\in[1,\infty)$.
Then the symplectic eigenvalues $(d_1,\dots,d_n)$ of the
matrix $\gamma$ satisfy
\begin{equation}\label{Cond2}
    \sum_{k=1}^n d_k \leq \sum_{k=1}^n c_k.
\end{equation}
\end{lemma}

{\it Proof:} Note that the right hand side of Eq.\ (\ref{Cond2}) is
nothing but half the trace of the covariance matrix
$\gamma$,
whereas the left hand side is the {\it symplectic trace}
$\text{str}(\gamma)$
of  $\gamma$, so
\begin{equation}
    \text{str}(\gamma) = \sum_{j=1}^n d_j
\end{equation}
if $\text{sspec}(\gamma) = (d_1,\dots, d_n)$,
see, e.g., Ref.\ \cite{Hyllus}.
We arrive at
this relationship by making use of a property of the
trace-norm.  The symplectic eigenvalues $d_1,\dots,d_n$
of $\gamma$ are given by the
 square roots of the simply counted
 eigenvalues of the matrix
  $ (i \sigma)  \gamma(i \sigma)
  \gamma$ \cite{Survey,Peter}.
  Hence, the symplectic spectrum is just given by the
  spectrum of the matrix
  \begin{equation}
    M= |\gamma^{1/2} (i \sigma) \gamma^{1/2}| ,
\end{equation}
where $| \cdot |$ denotes the matrix absolute 
value \cite{MAV}.
So
we have that
\begin{eqnarray}
    2
    \sum_{k=1}^n d_k &=&
    \text{tr}(M)
    =
    \|  \gamma^{1/2} (i \sigma) \gamma^{1/2}\|_1,
\end{eqnarray}
where $\| \cdot \|_1$ is the trace norm. The property we
wish to prove then immediately follows from the fact that the
trace-norm is a unitarily invariant norm: this implies that
\begin{eqnarray}
    2 \sum_{k=1}^n d_k =
    \|  \gamma^{1/2} (i \sigma) \gamma^{1/2}\|_1
    \leq \|(i\sigma) \gamma\|_1 ,
\end{eqnarray}
as $\|AB\|_1 \leq \|B A\|_1$ for any matrices
$A$, $B$ for which
$AB$ is   Hermitian. This
inequality holds for any unitarily invariant norm whenever
$AB$ is a normal operator \cite{Bhatia}.
Now, since any covariance
matrix is positive, $\gamma\geq 0$, and the
largest singular value of $i\sigma$ is clearly given by
unity, we can finally
conclude that
\begin{eqnarray}
    2 \sum_{k=1}^n d_k  \leq
    \|  \gamma \|_1 = \text{tr}(\gamma) =2  \sum_{k=1}^n c_k,
\end{eqnarray}
which is the statement that we intended to show.
\proofend

This observation implies as an immediate
consequence a necessary condition for the
possible reductions, given a Gaussian state of an
$n$-mode system: Let $\gamma$ be the covariance matrix of a
Gaussian pure state of $n$ modes, with reductions as above.
We can think of the state as a bi-partite
state between a distinguished mode labeled $k$, without
loss of generality being the last mode $k=n$, and the rest
of the system.
We can in fact Schmidt decompose this pure state with
respect to this split using Gaussian unitary operations
\cite{Giedke,Schmidt1,Holevo}. This means that we can find
symplectic transformations
$S_A\in \text{Sp}(2(n-1),\rr)$ and
$S_B\in \text{Sp}(2,\rr)$ such that
\begin{equation}
    (S_A\oplus S_B) \gamma
    (S_A \oplus S_B)^T =
    \left[
    \begin{array}{cc}
    A & C\\
    C^T & B\\
    \end{array}
    \right],
\end{equation}
where
\begin{eqnarray}
    A &=& \text{diag}(1,\dots,1 ,a_n,a_n),\\
    B &=& \text{diag}(a_n,a_n),
\end{eqnarray}
with some $2(n-1)\times 2$-matrix $C$.
The symplectic eigenvalues of modes $1,\dots , n-1$
are hence given by  $1,\dots, 1,a_n$. The above statement
therefore implies the inequality
\begin{equation}
    n-2 + a_n \leq a_1+ \dots + a_{n-1},
\end{equation}
or, by substituting $b_k=a_k-1$ for all $k=1,\dots ,n$,
\begin{equation}
    b_n \leq b_1+b_2+\dots + b_{n-1}.
\end{equation}
This must obviously hold for all distinguished modes and not
only the last one, and hence,
we arrive at the following simple necessary conditions:

\begin{corollary}[Necessary conditions for
pure states]\label{Necessary}
    Let $\gamma$ be the covariance matrix of
    a pure Gaussian state with thermal
    reductions
    \begin{equation}\label{The}
    \gamma_k=\left[
    \begin{array}{cc}
    b_k +1& 0 \\
    0 & b_k+1
    \end{array}
    \right],
\end{equation}
$k=1,\dots ,n$. Then, for all $j$,
\begin{equation}\label{Sim}
    b_{j} \leq \sum_{k\neq j} b_k.
\end{equation}
\end{corollary}
That is, the largest value of $b_j$ cannot exceed the sum of
all the other ones.  

So far, we have assumed the global state $\rho$ to be a pure state.
In the full problem, however, we may of course allow
$\rho$ to be any
Gaussian state, and hence a mixed one, with symplectic spectrum
\begin{equation}
    \text{sspec}(\rho) = (d_1,\dots, d_n)\geq (1,\dots, 1),
\end{equation}
instead of being $(1,\dots, 1)$. This is the Gaussian analogue of
the  mixed marginal problem. For this mixed state case, we
provide necessary conditions for the main reductions, in form
of $n$ inequalities on partial sums, and one where the largest
symplectic eigenvalue of a reduction plays an important role.
The first set of $n$ conditions is up to the different ordering
a weak majorization relation for
symplectic eigenvalues, which is in fact essentially
a corollary of a
result from Ref.\ \cite{Hir} due to Hiroshima.
The second statement, the
$n+1$-th condition, as well as showing sufficiency of the
general conditions, will turn out to be significantly more involved.

\begin{lemma}[Necessity of the first $n$ conditions]\label{Majorization}
    Let $(d_1,\dots, d_n)$,
    and $(c_1,\dots, c_n)$ be defined
    as in Theorem \ref{MixedProblem}. For any given
    $(d_1,\dots, d_n)$, the admissible $(c_1,\dots, c_n)$
    satisfy
    \begin{equation}\label{PS}
        \sum_{j=1}^k c_j
        \geq \sum_{j=1}^k d_j
    \end{equation}
    for all $k=1,\dots, n$.
\end{lemma}

{\it Proof:} Let $S\in \text{Sp}(2n,\rr)$ be the matrix
from the symplectic group
that brings $\gamma$ into
diagonal form, so
\begin{equation}\label{Williamson}
    S\gamma S^T = \text{diag}(d_1,d_1,\dots, d_n,d_n).
\end{equation}
The main diagonal elements of $\gamma$, in turn,
again without loss of generality in non-decreasing order,
are given by $(c_1,\dots, c_n)$.  Now according to Ref.\
\cite{Hir}, we have that
\begin{equation}\label{TheCond}
    \min
    \text{tr}(T \gamma T^T) = 2\sum_{j=1}^k d_j
\end{equation}
for $k=1,\dots, n$, where the minimum is taken over all real
$2k\times 2n$-matrices
$T$ for which
\begin{equation}\label{SF}
    T\sigma_n T^T = \sigma_k.
\end{equation}
Here, $\sigma_k$
denotes the symplectic matrix on $k$ modes as defined
in Eq.\ (\ref{SymplecticMatrix}).
Now we can actually take
$S\in \text{Sp}(2n,\rr) $
according to
    $S =
    \id $,
we see that $T$, consisting of the first $2k$ rows of $S$,
satisfies Eq.\ (\ref{SF}). Since this submatrix
does not necessarily correspond to a minimum in
Eq.\ (\ref{TheCond}), we find
\begin{eqnarray}
    2\sum_{j=1}^k c_j = \text{tr}(T \gamma T^T)
    \geq  2\sum_{j=1}^k d_j ,
\end{eqnarray}
for any $k=1,\dots, n$.
\proofend

We will now prove the necessity of the $n+1$-th
inequality constraint in Theorem \ref{MixedProblem}.

\begin{lemma}[Necessity of the last condition] \label{FullLemma}
    Let $(d_1,\dots, d_n)$,
    and $(c_1,\dots, c_n)$ be defined
    as in Theorem \ref{MixedProblem}. For any given
    vector of symplectic eigenvalues
    $(d_1,\dots, d_n)$, the admissible
    $(c_1,\dots, c_n)$ satisfy
    \begin{equation}\label{Tha}
    c_n  -  \sum_{j=1}^{n-1} c_j \leq
    d_n -
    \sum_{j=1}^{n-1} d_j  .
    \end{equation}
\end{lemma}

{\it Proof:} We will define the function
$f:{\cal S}_n\rightarrow\rr$, where ${\cal S}_n$ is
the set of strictly positive real $2n\times 2n$-matrices,
as follows: We define the
vector $c=(c_1,\dots, c_n)$ as
\begin{equation}
    c_j = (\gamma_{2j-1,2j-1} \gamma_{2j,2j} - \gamma_{2j-1,2j}^2 )^{1/2}
\end{equation}
$j=1,\dots, n$,
as the usual vector of symplectic
spectra of each of the $n$ modes,
and then set
\begin{equation}
    f(\gamma) :=  2 \max(c) - \sum_{j=1}^n c_j.
\end{equation}
For a diagonal matrix
    $D=\text{diag}(d_1,d_1,\dots, d_n,d_n)$
with entries in non-decreasing order,
we have
\begin{equation}\label{Before}
    f(D) = d_n-\sum_{j=1}^{n-1} d_j.
\end{equation}

We will now investigate the
orbit of this function $f$ under the
symplectic group,
\begin{equation}
    \tilde f =\sup \left\{x\in \rr: x= f(SDS^T), \, S\in \text{Sp}(2n,\rr)\right\},
\end{equation}
and will see that the supremum is actually attained
as a maximum for $S=\id$. Each of the
matrices $\gamma= SDS^T$ have by construction the same
symplectic spectrum as $D$. This is a variation over
$2n^2+n$ real parameters, as any
$S\in \text{Sp}(2n,\rr) $ can be decomposed
according to the {\it Euler decomposition} as
\begin{equation}
    S= O Q V,
\end{equation}
where $O,V\in K(n):= \text{Sp}(2n,\rr)\cap O(2n)$ and
\begin{equation}
    Q\in \left\{
    (z_1,1/z_1,\dots, z_n,1/z_n):
    z_k \in \rr\backslash\{0\}
    \right\}.
\end{equation}
That is, $O,V$ reflect passive operations,
whereas $Q$ stands for a squeezing operation.

We will now see that the maximum of this function
$f$ -- which exists, albeit the group being non-compact --
is actually attained when the matrix is already diagonal. This means that in general, we have
that
\begin{equation}
    \tilde f =  2 \max \text{sspec}(\gamma) - \text{str}(\gamma).
\end{equation}
For any global maximum, any local variation will not increase
this function further. Let us start from
some $\gamma = SDS^T$. For any such covariance
matrix $\gamma$ we can find a $T\in \text{Sp}(2(n-1),\rr)$
such that
\begin{equation}\label{SimpleForm}
    (T\oplus \id_2)\gamma (T\oplus \id_2)^T =
    \left[
    \begin{array}{cc}
    E & F\\
    F^T & G
    \end{array}
    \right] =:\gamma',
\end{equation}
where
\begin{equation}
    E=\text{diag}(c_1',c_1',\dots, c_{n-1}',c_{n-1}')
\end{equation}
is a $(2n-2)\times (2n-2)$ matrix and $G$ is a
$2\times 2$ matrix. Using Lemma \ref{ST} again, we find that
\begin{equation}
    \sum_{j=1}^{n-1} c_j' \leq \sum_{j=1}^{n-1} c_j,
\end{equation}
so
\begin{equation}
    f(\gamma')\geq f(\gamma).
\end{equation}
In other words, it does not restrict generality to assume the
final covariance matrix to be of the form as in the right hand
side of Eq.\ (\ref{SimpleForm}), and we will use the notation
\begin{equation}\label{Prepa2}
    \gamma = S D S^T = \left[
    \begin{array}{cc}
    E & F\\
    F^T & G
    \end{array}
    \right]
\end{equation}
with $E=\text{diag}(c_1,c_1,\dots, c_{n-1},c_{n-1})$
and $G=\text{diag}(c_{n},c_{n})$.

We can now investigate submatrices of $\gamma$ associated with
modes $m$ and $n$, $1\leq m<n$,
\begin{equation}
    M_{m,n}= \left[
    \begin{array}{cc}
    c_m \id_2 & C_{n,m}\\
    C^T_{n,m} & c_n \id_2
    \end{array}
    \right].
\end{equation}
This we can always bring to a diagonal form, using symplectic diagonalization, only affecting
the main diagonal elements of modes $n$ and $m$, and leaving the other main diagonal
elements invariant. This brings this submatrix into the form
\begin{equation}
    M_{m,n}'= \left[
    \begin{array}{cc}
    c_m' \id_2 & 0\\
    0 & c_n' \id_2
    \end{array}
    \right],
\end{equation}
with $c_n'\geq c_m'$. From Lemma \ref{Difference} we know that
\begin{equation}
    c_n' - c_m' \leq  c_n - c_m,
\end{equation}
so we have increased the function $f$, whenever
$C_{n,m}\neq 0$. Hence, for global and hence local
optimality, we have to have $C_{n,m}=0$. However, each of the matrices $C_{n,m}=0$
for all $m=1,\dots, n-1$ exactly
if the matrix $\gamma$ is already diagonal.

What remains to be shown is that the function $f$ is bounded
from above, to exclude the
case that the maximum does not even exist. One way to
show this is to make use of the upper bound in Lemma
\ref{FromPure} to have
for every covariance matrix $\gamma$ with
symplectic spectrum $(d_1,\dots, d_n)$
in non-decreasing order
\begin{equation}
    f(\gamma) \leq   \sum_{j=2}^n d_j  +(3-2 n) d_1,
\end{equation}
which shows that $f$ is always bounded from above. If
$\gamma$ is merely a strictly positive real matrix, but no
covariance matrix, an upper bound follows from a
rescaling with a positive number.
\proofend

%


We now prove the upper bound required for the proof of Lemma \ref{Necessary}.

\begin{lemma}[Upper bound]\label{FromPure}
    Let $(d_1,\dots, d_n)$,
    and $(c_1,\dots, c_n)$ be defined
    as in Theorem \ref{MixedProblem}, and $\gamma$
    be additionally a $2n\times 2n$ covariance matrix.
    For any given
    $(d_1,\dots, d_n)$, the admissible $(c_1,\dots, c_n)$
    satisfy
    \begin{equation}\label{PureConditionsMixed}
    c_n  -  \sum_{j=1}^{n-1} c_j \leq
    \sum_{j=2}^n d_j  +(3-2 n) d_1,
    \end{equation}
\end{lemma}

{\it Proof:}
We start from a $4n\times 4n$-covariance matrix
\begin{equation}\label{OCM}
    \gamma =\left[
    \begin{array}{cc}
    A & C\\
    C^T & A
    \end{array}
    \right] ,
\end{equation}
corresponding to a pure Gaussian state,
where
\begin{eqnarray}
    A &=& \bigoplus_{k=1}^n
    \left[
    \begin{array}{cc}
    d_{k} & 0\\
    0 & d_{k}
    \end{array}
    \right],\\
    C&=& \bigoplus_{k=1}^n
    \left[
    \begin{array}{cc}
    (d_{k}^2-1)^{1/2} & 0\\
    0 & -(d_{k}^2-1)^{1/2}
    \end{array}
    \right]
\end{eqnarray}
are real $2n\times 2n$-matrices. Physically,
this means that we start from a
collection of $n$ two mode squeezed
states, with the property that the reduction to the first $n$
modes is just a diagonal covariance matrix with symplectic
eigenvalues $(d_1,\dots,d_n)$, again in non-decreasing order.
Let us first assume that $d_1=1$, this assumption will be relaxed
later. Let us now consider
\begin{equation}\label{Tr}
    \left[
    \begin{array}{cc}
    S_1 & 0\\
    0 & \id
    \end{array}
    \right]\gamma
        \left[
    \begin{array}{cc}
    S_1^T & 0\\
    0 & \id
    \end{array}
    \right] = \left[
    \begin{array}{cc}
    S_1AS^T_1 & S_1 C\\
    C^T S_1^T& A
    \end{array}
    \right],
\end{equation}
for $S_1\in \text{Sp}(2n,\rr)$. Obviously, the
set we seek to characterize is the set ${\cal B}$
of main diagonals
of the upper left block
\begin{equation}
    U=S_1AS^T_1
\end{equation}
of this matrix. We can always
start from a diagonal matrix having the symplectic
eigenvalues on the main diagonal, and consider the
orbit under all symplectic transformations $S\in \text{Sp}(4n,\rr)$.

We will now relax the problem by allowing all
$S\in \text{Sp}(4n,\rr)$ instead of
symplectic transformations of the form $S=S_1\oplus\id$,
$S_1\in \text{Sp}(2n,\rr)$. We hence consider the
full orbit under all symplectic transformations.
This set  ${\cal C} \supset {\cal B}$
is characterized by the
reductions to single modes of
 \begin{equation}\label{BS}
    \gamma' =
    S\gamma S^T = \left[
    \begin{array}{cc}
    A' & C'\\
    {C'}^T & A
    \end{array}
    \right]
\end{equation}
for some $S\in \text{Sp}(4n,\rr)$, such that again
    $A= \text{diag}(d_1,d_1,\dots, d_n,d_n)$
This includes
the case (\ref{Tr}).

We are now in the position to make use of
the statement that we have established
before: From exploiting the Schmidt decomposition on
the level of second moments, and using Lemma \ref{ST}
relating the trace to the symplectic trace,
we find
\begin{equation}\label{Full}
    c_n  -  \sum_{j=1}^{n-1} c_j \leq
    \sum_{j=2}^n d_j  +3-2 n,
\end{equation}
as $d_1=1$ was
assumed.

Let us now consider the case of $d_1>1$. We will apply
the previous result, after appropriately rescaling the covariance matrix.
Indeed, we
can construct a covariance matrix $\tilde\gamma$
as in Eq.\ (\ref{OCM}) for
\begin{equation}
    (\tilde d_1,\dots, \tilde d_n)
    = (1,d_2/d_1,\dots, d_n/d_1).
\end{equation}
We then investigate
the orbit of
    $\tilde \gamma $
under the symplectic group, and
look at the main diagonal elements of
    $S\tilde \gamma S^T $.
By construction, we have that
$\tilde \gamma+i\sigma\geq 0$. We can hence
apply Eq.\ (\ref{Full}) to this case.
Multiplying both sides of Eq.\ (\ref{BS}) by $d_1$
gives rise to the condition in Eq.\ (\ref{PureConditionsMixed}).
\proofend

\begin{lemma}[Solution to two-mode problem]\label{Difference}
There exists a strictly positive real $4\times4$-matrix
$\gamma$ with main diagonal blocks
$\text{diag}(c_1,c_1),\text{diag}(c_2,c_2)$
and symplectic eigenvalues $(d_1,d_2)$ if and only
if
\begin{eqnarray}
    c_1+c_2 &\geq& d_1+d_2,\\
    c_2-c_1 &\leq & d_2-d_1,
\end{eqnarray}
assuming $c_2\geq c_1$ and $d_2\geq d_1$.
Moreover, $c_1-c_2 = d_1-d_2$ if and only
the $2\times 2$ off diagonal
block of $\gamma$ vanishes.
\end{lemma}

{\it Proof:} The necessary conditions that $|c_1-c_2|\leq |d_1-d_2|$
are a consequence of Lemma \ref{FromPure}. The
necessary conditions
$c_1+c_2\geq d_1+d_2$ and $c_1\geq d_1$
have been previously shown in
Lemma \ref{Majorization}. Hence, we have to show that these
conditions can in fact be achieved. This can be done
by considering a
\begin{equation}
    \gamma=\left[
    \begin{array}{cccc}
    c_1 & 0 &e & 0 \\
    0 & c_1 & 0 & f \\
    e & 0 & c_2 & 0\\
    0 & f & 0&  c_2
    \end{array}
    \right] = S\text{diag}(d_1,d_1,d_2,d_2)S^T.
\end{equation}
The relationship between $c_1,c_2,e,f$ and $d_1,d_2$
is given by
\begin{eqnarray}
    d_{1/2}^2 &= &\bigl(c_1^2 + c_2^2+ 2 ef
    \nonumber\\
    &\pm&
    (c_1^4 + c_2^4 + 4 ef c_2^2  - 2 c_1^2 (c_2^2 - 2 ef)
    + 4
    c_1 c_2 (e^2 + f^2))^{1/2}\bigr)/2,
\end{eqnarray}
as $d_1,d_2$ are the square roots of the
eigenvalues of $-\sigma\gamma\sigma\gamma$ \cite{Survey}, 
compare also Ref.\ \cite{Twomodes}.
An elementary analysis shows that the above
inequalities can always be achieved. Also, the extremal values
are achieved if and only if $e=f=0$.
\proofend

What we finally need to show is that the conditions that we have
derived are in fact sufficient.  This will be the most involved
statement.

\begin{lemma}[Sufficiency of the conditions]
For any vectors $(c_1,\dots, c_n)$ and $(d_1,\dots, d_n)$
satisfying the conditions (\ref{C1}) and (\ref{C2}) there
exists a $2n\times 2n$ strictly positive real matrix with
$\text{diag}(c_1,\dots, c_n)$ as its symplectic
main diagonal
elements and $(d_1,\dots, d_n)$ as its symplectic eigenvalues.
\end{lemma}

{\it Proof:} The argument will essentially be an argument by
induction, in several ways resembling the argument put forth
in Refs.\ \cite{Sing,Tho1,Tho}. The underlying
idea of the proof is essentially as follows: On using the
given constraints, one constructs an appropriate
$2(n-1)\times 2(n-1)$-matrix, in a
way that it can be combined to the
desired $2n\times 2n$-matrix by means of an
appropriate $S\in \text{Sp}(4,\rr)$ acting on a $4\times4$
submatrix only. Note, however, that
compared to Ref.\ \cite{Tho},
we look at
variations over the non-compact symplectic group $\text{Sp}(2n,\rr)$,
and not the compact $U(2n)$.

For a single mode, $n=1$, there
is nothing to be shown. For two modes, Lemma \ref{Difference}
provides the sufficiency of the conditions. Let us hence assume
that we are given vectors $(c_1,\dots, c_n)$
and $(d_1,\dots, d_n)$ as above, and that we have already
shown that for $2(n-1)\times 2(n-1)$-matrices,
the conditions (\ref{C1}) and (\ref{C2})
are indeed sufficient. We complete the proof by explicitly
constructing an $2n\times 2n$-matrix with the stated property.

We have that $c_1\geq d_1$ by assumption. We could also
have $c_1\geq d_j$ for some $2\leq j\leq n$, so let
$k\in\{1,\dots,n\}$
be the largest index such that
\begin{equation}
    c_1\geq d_k.
\end{equation}
Let us first consider the case that $k\leq n-2$, and we will
consider the cases $k=n-1$ and $k=n$ later.
Then we can set
$x:= d_k + d_{k+1}- c_1$, which means
that $x\geq 0$, and that all conditions
\begin{eqnarray}
    c_1 + x &\geq& d_k + d_{k+1},\label{c1}\\
    c_1 - x &\geq& d_k - d_{k+1},\label{c2}\\
    - c_1 + x &\geq& d_k - d_{k+1}\label{c3}
\end{eqnarray}
are satisfied: (\ref{c1}) by definition, (\ref{c2}) because
$c_1\geq d_k$ and (\ref{c3}) as $d_{k+1}\geq c_1$.
This means that we can find a matrix
of the form
\begin{eqnarray}
    \gamma' &:=&
    \left[
    \begin{array}{cc}
    c_1 \id_2 & C\\
    C^T & x\id_2
    \end{array}
    \right],
\end{eqnarray}
for some $2\times 2$-matrix $C$, with symplectic eigenvalues
$(d_k,d_{k+1})$, using Lemma \ref{Difference}. Therefore, the
matrix
\begin{eqnarray}
    \gamma'' =
    \gamma'
    \oplus
    \text{diag}(d_1,d_1,d_2,d_2,\dots, d_{k-1},d_{k-1},
     d_{k+2},d_{k+2},\dots, d_n,d_n)
\end{eqnarray}
has the symplectic spectrum
$(d_1,\dots, d_n)$.

We will now show that we can
construct a $2(n-1)\times 2(n-1)$ matrix $\gamma'''$
with symplectic eigenvalues $(d_1,\dots, d_{k-1},x,d_{k+2},
\dots, d_n)$ and main diagonal elements
$(c_2,c_2,\dots, c_n,c_n)$, by invoking the induction
assumption. This matrix $\gamma'''$
we can indeed construct, as we have
\begin{eqnarray}
    c_2+ \dots + c_l &\geq& d_1 +\dots + d_{l-1},\,l=2,\dots, k,\\
    c_2+ \dots + c_{k+1} &\geq& d_1 +\dots + d_{k-1}+ x,\\
    c_2+ \dots + c_{s} &\geq& d_1 +\dots + d_{k-1}+ x
    +
    d_{k+2} + \dots + d_{s}, s=k+2,\dots, n, 
\end{eqnarray}
as one can show using $d_k\leq c_1 \leq d_{k+1}$ and
$x=d_k + d_{k+1}- c_1$. Also, we have
\begin{equation}
    c_n - c_2- \dots  - c_{n-1}\leq d_n
    - d_1- \dots - d_{k-1} - x  -
    d_{k+2} - \dots - d_n,
\end{equation}
fulfilling all of the condition that we need invoking the
induction assumption to construct $\gamma'''$. This
matrix has the same symplectic eigenvalues as the
right lower $2(n-1)\times 2(n-1)$ submatrix $\gamma''''$
of $\gamma''$.
Therefore, there exists an $S\in \text{Sp}(2(n-1),\rr)$ such that
\begin{equation}
    \gamma'''' = S\gamma''' S^T.
\end{equation}
So the matrix
\begin{equation}
    \gamma := (\id_2\oplus S) \gamma'' (\id_2\oplus S)^T
\end{equation}
has the symplectic spectrum $(d_1,\dots, d_n)$
and symplectic
main diagonal elements $(c_1, \dots, c_n )$.
Hence, by invoking the induction assumption, we have been
able to construct the desired matrix with the appropriate
symplectic spectrum and main diagonal elements. Note that
only two-mode operations have been needed in order to achieve
this goal.

We now turn to the two remaining cases, $k=n$ and $k=n-1$.
In both cases this means that we have
$c_1\geq d_{n-1}$, as $d_n\geq d_{n-1}$, and both
cases can be treated in actually exactly the same manner.
Obviously, this implies that also
$c_n\geq c_1 \geq d_{n-1}$. We can now define again an
$x$, by means of a set of inequalities. This construction is
very similar to the one in Ref.\ \cite{Tho}.
We can require on the one hand
\begin{eqnarray}
    x &\geq&  \max \{ d_{n-1}, d_{n-1} + d_n - c_n,
    d_{n-1} - d_n + c_n,\nonumber\\
    &&d_1+ \dots + d_{n-2} + c_{n-1}
    - c_1 - \dots - c_{n-2}\}.
\end{eqnarray}
On the other hand, we can require
\begin{eqnarray}
    x &\leq&  \min \{  d_n - d_{n-1} + c_n,
    c_1+ \dots + c_{n-1} - d_1 - \dots- d_{n-2},
    0\}.
\end{eqnarray}
Both these conditions can be simultaneously satisfied, making
use of $c_n \geq c_{n-1}$ and $c_n\geq d_{n-1}$.
This in turn means that we have
\begin{eqnarray}
    c_n + x \geq d_{n-1} + d_n,\\
    c_n - x \geq  d_{n-1} - d_n,\\
    x - c_n \geq  d_{n-1} - d_n,
\end{eqnarray}
where the latter two inequalities mean that
$|x -c_n| \leq |d_{n-1} - d_n|$.
Moreover, we satisfy all the inequalities
\begin{eqnarray}
    c_1+ \dots + c_l \geq c_1+ \dots + d_l,\, l=1,\dots, n-2,\\
    c_1 + \dots + c_{n-1} \geq d_1 + \dots + d_{n-2} +x,
\end{eqnarray}
and
\begin{equation}
    c_{n-1} - c_1 - \dots - c_{n-2}
    \leq x - d_1 - \dots - d_{n-2}.
\end{equation}
Again, we can hence invoke the induction assumption, and
construct in the same way as before the desired
covariance matrix with symplectic spectrum $(d_1,\dots, d_n)$
and symplectic main diagonal elements $(c_1,\dots, c_n)$.
This ends the proof of sufficiency of the given conditions.
\proofend

\section{Physical implications of the result and outlook}

The results found in this work can also be read as a full
specification of what multipartite Gaussian states may be
prepared: Since the argument is constructive it readily provides
a recipe of how to construct {\it multi-mode Gaussian entangled
states} with all possible local entropies: For pure states, starting from
squeezed modes, all is needed is a network of passive operations.
Applied to optical systems of several modes,
notably, this gives rise to
a protocol to prepare multi-mode pure-state entangled light of
all possible entanglement structures from squeezed light, using
passive linear optical networks, via
\begin{equation}
    \gamma = OPO^T,
\end{equation}
with $P= (z_1,1/z_1,\dots, z_n,1/z_n)$, $z_k\in \rr\backslash\{0\}$,
and $O\in K(n)$. $P$ is the covariance matrix of squeezed
single modes, whereas $O$ represents the passive optical
network. The latter can readily be broken down to
a network of beam splitters and phase
shifters, according to Ref.\  \cite{Reck}.
Hence, our result also generalizes
the preparation of Ref.\ \cite{Alessio} from
the case of three modes
to any number of modes. Similarly, for mixed states, the given
result readily defines a preparation procedure, but now using also squeezers in general.

The above statement also settles the
question of the {\it sharing of entanglement} 
of single modes versus
the rest of the system in a multi-mode system: 
For a pure Gaussian state
with $d_1=\dots = d_n=1$, the 
entanglement entropy 
$E_{j|\{1,\dots, n\}\backslash \{j\}}$
of a mode labeled $j$ with respect to the rest
of the system is given by
\begin{eqnarray}\label{edef}
	E_{j|\{1,\dots, n\}\backslash \{j\}}:=
	S(\rho_j) = s(c_j):= \frac{c_j+1}{2}\log_2 \frac{c_j+1}{2}
	-  \frac{c_j-1}{2}\log_2 \frac{c_j-1}{2},
\end{eqnarray}
where $s:[1,\infty)\rightarrow [0,\infty)$ is a monotone
increasing, concave function. 

\begin{corollary}[Entanglement sharing in pure Gaussian states] 
For pure Gaussian states, the set of all 
possible entanglement
values of a single 
mode with respect to the system is given by
\begin{equation}
	\left(
	E_{1 |\{2,\dots, n\} },\dots,
	E_{n|\{1,\dots, n-1\}
	}
	\right) \in
	\left\{
	(s(c_1),\dots, s(c_n)):
	c_j-1\leq\sum_{k\neq j}
	(c_k -1 ),\,c_j\geq 1
	\right\}.
\end{equation}
\end{corollary}

This result is an immediate consequence of the above 
pure marginal problem, Corollary \ref{pureprob}. 
In fact, this is for pure Gaussian states
more than a monogamy inequality: it constitutes a full
characterization 
of the complete set of consistent degrees of entanglement.

A further practically useful application of our result is
the following: It tells us how {\it pure} a state must have been,
based on the information available from 
{\it measuring local properties} like local
photon numbers. This is expected to be a very desirable 
tool in an experimental context: In optical systems, such
measurements are readily available with homodyne 
or photon counting measurements. 

\begin{corollary}[Locally measuring global purity in non-Gaussian states] Let us assume
that one has acquired knowledge about the local
symplectic eigenvalues $c_1,\dots, c_n$ of a global state
$\rho$. Then one can infer
that the global von-Neumann entropy $S(\rho)$ of $\rho$
satisfies
\begin{equation}
	S(\rho)\leq s\biggl(\sum_{k=1}^n c_k\biggr).
\end{equation}
This estimate is true regardless whether the state
$\rho$ is a Gaussian state or not.
\end{corollary}

{\it Proof:} Let us denote with $\omega$ the Gaussian
state with the same covariance matrix $\gamma\geq 0$
as the (unknown) state $\rho$. The vectors 
$(d_1,\dots, d_n)$ and $(c_1,\dots, c_n)$ are the
symplectic eigenvalues and symplectic main diagonal
elements of $\omega$, respectively. From the fact that
$\text{diag}(d_1,d_1,\dots,d_n,d_n)$ reflects a tensor
product of Gaussian states, we can conclude that
\begin{equation}
	S(\omega)=\sum_{j=1}^n s(d_j).
\end{equation}
In turn, from Lemma \ref{ST} we find that
\begin{equation}
	\sum_{j=1}^n c_j \geq \sum_{j=1}^n d_j.
\end{equation}
By means of an extremality property of the von-Neumann
entropy (see, e.g., Ref.\
\cite{Holevo,Channels}) that a Gaussian state has the
largest von-Neumann entropy for fixed second moments, 
we find that $S(\rho) \leq S(\omega)$. 
Since the  function $s:[1,\infty)\rightarrow [0,\infty)$
defined in Eq.\ (\ref{edef}) is concave and monotone increasing, 
we have that
\begin{eqnarray}
	S(\rho)\leq \sum_{j=1}^n s(d_j)\leq s(d_1+\dots+d_n)
	\leq s(c_1+\dots+c_n).
\end{eqnarray}
This is the statement that we intended to prove. Clearly, this bound is 
tight, as is obvious when applying the inequality to the 
Gaussian state with covariance matrix 
$\text{diag}(d_1,d_1,\dots, d_n,d_n)$ itself.
\proofend

For example, if obtains
$c_1=3/2=c_2=3/2$ and $c_3=2$ in local measurements
on the local photon number, then
one finds that the global state
necessarily satisfies $S(\rho)\leq s(5)$. This is a powerful
tool when local measurements in optical systems
are more accessible than
global ones, for example, when no phase reference is 
available, or bringing modes together is a difficult task.

To finally turn to the role of Gaussian operations in this work:
Our result 
highlights an observation that has been encountered already
a number of times in the literature: That global Gaussian
operations applied to many modes at once
are often hardly more powerful than when applied to
pairs of modes. This resembles to some extent the situation in
the distillation of entangled Gaussian states by means of
Gaussian operations  \cite{Survey,Op,Fiurasek,GiedkeOperations}.

In this work, we have given a complete characterization of
reductions of pure or mixed Gaussian states. In this way,
we have also given a general picture of the possibility
of sharing quantum correlations in a continuous-variable
setting. Since our proof is constructive, it also gives rise to a
general recipe to generate multi-mode entangled states with
all possible reductions. Formally, we established a connection
to a compatibility argument of symplectic spectra, by means
of new matrix inequalities fully characterizing the set in question.
These matrix inequalities formally resembles the well-known
Sing-Thompson Theorem relating singular values to main 
diagonal elements. It is the 
hope that this work can provide a significant insight into 
the achievable correlations
in composite quantum systems of many modes. 

\section{Acknowledgments}

We would like to thank V.\ Buzek,
P.\ Hyllus, and M.M.\ Wolf
for valuable
discussions on the subject of the paper, and especially
K.\ Audenaert for many constructive and helpful
comments concerning the
presentation of the results, and A.\ Serafini and G.\ Adesso
for further remarks on the manuscript. 
This work has been supported by the DFG
(SPP 1116, SPP 1078), the EPSRC, the QIP-IRC, iCORE,
CIAR,
Microsoft Research, and
the EURYI Award.

\end{document}